\documentclass{iaus}
\usepackage{graphicx}

\title[Statistical Properties of the Turbulent ISM] 
{Remarks on Statistical Properties of the Turbulent Interstellar Medium}

\author[M.A. de Avillez \& D. Breitschwerdt]   
{Miguel A. de Avillez$^{1,2}$ and Dieter Breitschwerdt$^2$}

\affiliation{$^1$Department of Mathematics, University of \'Evora, R.Rom\~ao 
Ramalho 59, 7000 \'Evora, Portugal
\break email:mavillez@galaxy.lca.uevora.pt\\[\affilskip] $^2$Institut f\"ur
Astronomie, Universit\"at Wien, T\"urkenschanzstr. 17, A-1180 Wien, Austria
\break email: breitschwerdt@astro.univie.ac.at}

\pubyear{2006}
\volume{237}  
\pagerange{119--126}
\date{?? and in revised form ??}
\jname{Triggered Star Formation in a Turbulent ISM}
\editors{B. G. Elmegreen \& J. Palous, eds.}
\begin{document}

\maketitle

\begin{abstract}
The supernova-driven interstellar medium in star-forming galaxies has Reynolds
numbers of the order of $10^{6}$ or even larger. We study, by means of adaptive
mesh refinement hydro- and magnetohydrodynamical simulations that cover the
full available range (from 10 kpc to sub-parsec) scales, the statistical
properties of the turbulent interstellar gas and the dimension of the most
dissipative structures. The scalings of the structure functions are consistent
with a log-Poisson statistics of supersonic turbulence where energy is
dissipated mainly through shocks.

\end{abstract}

\firstsection 
\section{Introduction}

The interstellar medium is a highly turbulent compressible flow with Reynolds
numbers of the order of $10^{6}$ or even larger (e.g., Elmegreen \& Scalo 2005).
Supernovae are the main source of energy of the interstellar gas, which
is dominated by nonlinear processes, including heating and cooling, that
transfer energy among a wide range of scales. It is unclear what are
the driving scales in the interstellar medium as observations show different
values ranging from a few parsec to tens or even hundreds of parsec (see, e.g.,
Haverkorn et al. 2005). Furthermore, it is still under debate which dissipation
process dominates the interstellar turbulence. High-resolution simulations of
the interstellar medium that cover a wide range of scales, including the
injection and viscous scales, are valuable tools to adress these issues.
We, therefore, carried out kpc-scale high resolution (up to 0.625 pc)
hydro- and magnetohydrodynamical simulations of the supernova-driven
interstellar medium (Avillez \& Breitschwerdt 2004, 2005, 2006) allowing us to
tackle simultaneously the disk and halo evolution and to understand the issues
referred above. In this paper we briefly discuss these results. Section 2 deals
with the injection scales, the scalings of the velocity structure functions and
dissipation processes are referred to in section 3, followed by a discussion of
the results and their implications (section 4).

\section{The Injection Scale of Interstellar Turbulence.}

The outer scale of the turbulent flow in the ISM is related to the scale at
which the energy in blast waves is transferred to the interstellar gas. Such a
scale can be determined by using the so-called two-point correlation function
$R_{ij}(\vec{l},t)=\langle u_{i}(\vec{x}+\vec{l},t)u_{j}(\vec{x},t)\rangle$,
where $u_{i}$ are the components of the fluctuating velocity field $\vec{u}$.
The diagonal components of $R_{ij}$ are even functions of $\vec{l}$ and can be
written in terms of dimensionless scalar functions $f(l,t)$ and $g(l,t)$,
which satisfy $f(0)=g(0)=1$ and $f,~g\leq 1$, that is, $R_{11}/u^{2}=f(l,t)$ and
$R_{ij}/u^{2}=g(l,t)$ if $i=j\neq 1$ and zero if $i\neq j$ with
$u=(\frac{1}{3}\langle \vec{u}\rangle)^{1/2}$. In these equations $R_{11}$
and $R_{22}=R_{33}$ are known as the longitudinal and transverse autocorrelation
functions, respectively.
\begin{figure}[thbp]
\centering
\includegraphics[width=0.32\hsize,angle=-90]{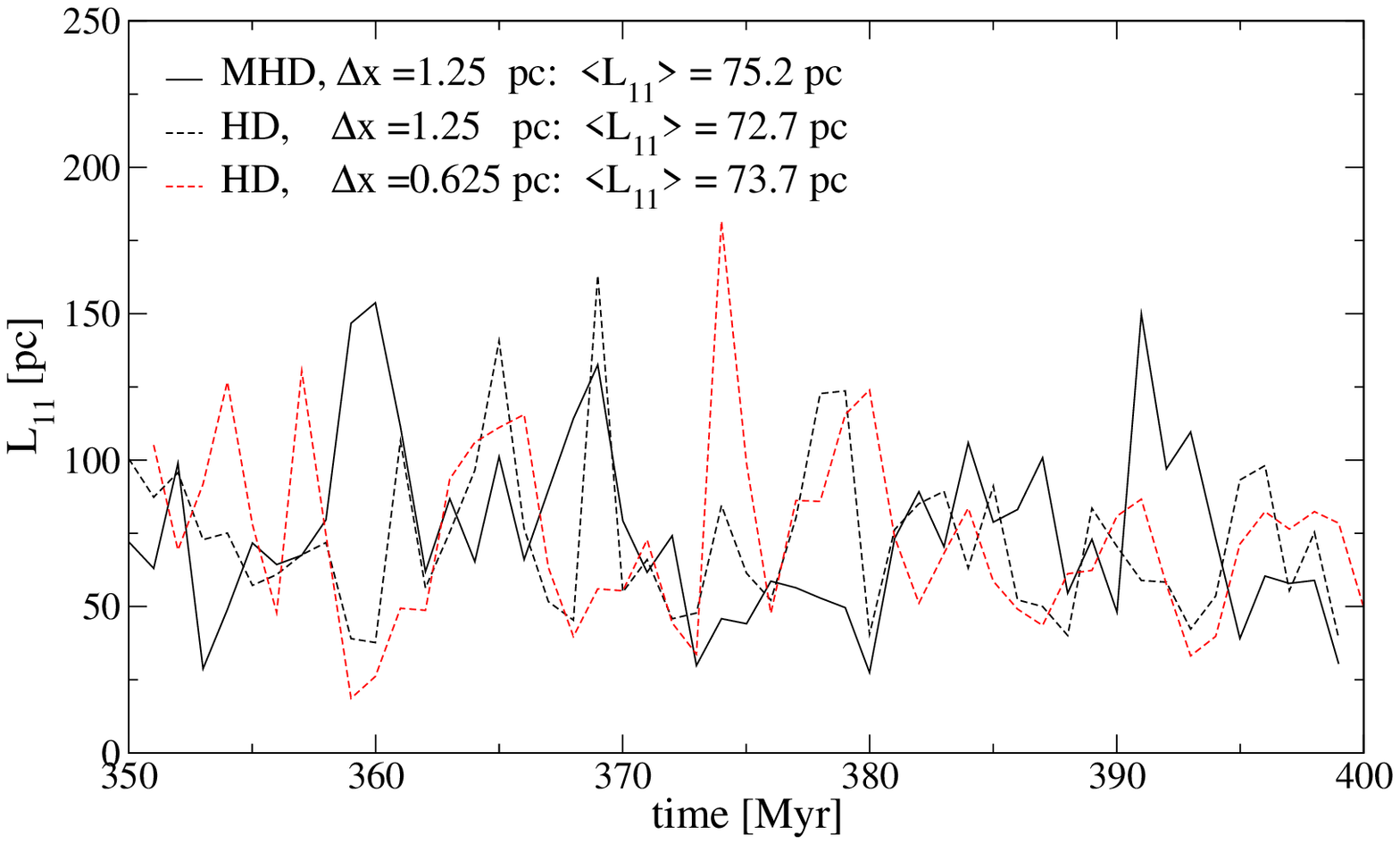}\includegraphics
[width=0.32\hsize,angle=-90]{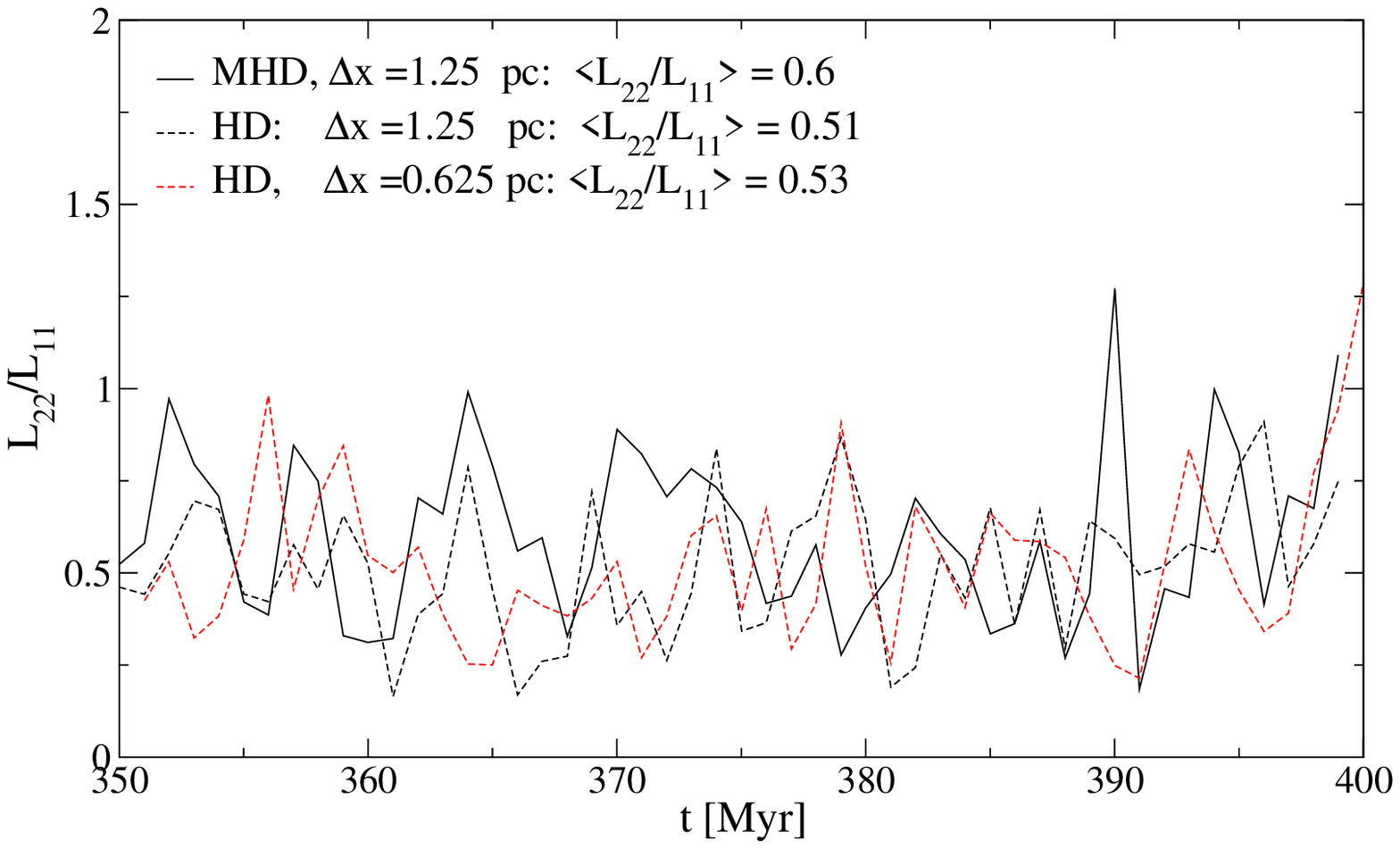}
\caption{History of the characteristic size (given by $L_{11}$) of
the larger eddies (left panel) and of the ratio $L_{22}/L_{11}$ (right
panel) for the HD (dashed lines) and MHD (solid line) runs for 1.25 pc
(black) and 0.625 pc (red) resolutions.}
\label{l11}
\end{figure}

The outer scale of the flow is given by $L_{11}=\int_{0}^{+\infty}f(l,t) dl$,
which is calculated in a region with a linear size of 500 pc at a distance of
250 pc from the edges of the computational domain such that we avoid the
periodicity effects of the
boundary conditions. The top panel of Figure~\ref{l11} shows the history
of $L_{11}$ in the last 50 Myr of evolution of the simulated interstellar medium
for 1.25 (in black) and 0.625 pc (in red) resolutions. The average integral
length scale in the three simulations is 73-75 pc and seems independent of the
resolution. In all the cases there is a large scatter of $L_{11}$ around its
mean as a result of oscillations in the local Galactic star formation rate,
where the formation and merging of superbubbles, is responsible
for the peaks observed in the two plots. The correlation length scale measured
in the simulations is remarkably similar to that determined by Kaplan (1958) who
found that interstellar velocities start to decorrelate at a scale of 80 pc.

The transverse integral length scale given by $L_{22}=\int_{0}^{+\infty}g(l,t)
dr$ equals $0.5 L_{11}$ in the case of isotropic turbulence. In the present
simulations we have $0.2<L_{22}/L_{11}<1.3$ (bottom panel of Figure~\ref{l11}).
In spite of the large scatter the time average of the $L_{22}/L_{11}$ over a
50 Myr period is 0.51 and 0.6 for the HD and MHD runs, respectively. The
discrepancy of $<L_{22}/L_{11}> \sim 0.5$ by about 20\% in the MHD run is a
consequence of the anisotropy in the flow introduced by the magnetic field.
Overall the field strength is still too low to produce a larger
deviation. In case of the HD run $<L_{22}/L_{11}>\sim 0.5$, indicates that in a
statistical sense the interstellar unmagnetised turbulence is roughly isotropic.

\section{Velocity Structure Functions and the Most Dissipative Structures}

Turbulent flows are usually described in a statistical fashion by
the velocity structure functions of order $p$ defined as
$S_{p}(l)=\langle \delta v_{l}^{p} \rangle \propto l^{\zeta(p)}$,
where $\zeta(p)$ is a scalar, $\delta v_{l}
=\left|v(x+l)-v(x)\right|$ with $v(x+l)$ and $v(x)$ being the
velocities along the $x-$axis at two points separated by a distance
$l$ such $\eta \ll l\ll L_{11}$, with $L_{11}$ being the energy
integral scale and $\eta=\nu^{3/4}\epsilon^{-1/4}$ the Kolmogorov
microscale, respectively. Here, $\nu$ is the kinematic viscosity and
$\langle\;\rangle$ stands for the ensemble average over the
probability density function of $\delta v_{l}$. From the Kolmogorov (1941;
hereafter denoted by K41) theory one derives
\begin{equation}
\displaystyle \langle \delta v_{l}^{p}\rangle\propto \langle \delta
v_{l}^{3}\rangle^{\zeta(p)/\zeta(3)}.
\end{equation}
with $l\gg \eta$. This property, referred to as extended self-similarity (ESS),
indicates that $S_{p}$ is dictated solely by the third-order structure function
$S_{3}$ via the set of scalings $\zeta(p)/\zeta(3)$ and is valid in a wide range
of length scales for large as well as small Reynolds numbers even if no inertial
range is established (Benzi et al. 1993). 
\begin{figure}[thbp]
\centering
\includegraphics[width=0.5\hsize,angle=-90]{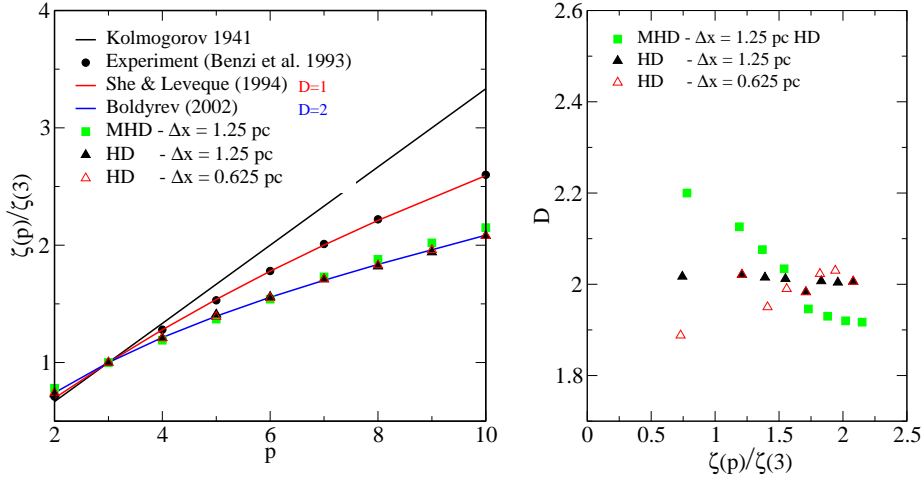}
\caption{\emph{Left:} Exponent $\zeta(p)/\zeta(3)$ for the structure function
versus order $p$. The full line corresponds to $\zeta(p)=p/3$ (K41 theory),
bullets correspond to data of Benzi et al. (1993), red and blue lines refer to
the She-Lev\'eque (1994) and Burgers-Kolmogorov (Boldyrev 2002) models,
respectively. Triangles represent data from HD simulations with 1.25 pc (black)
and 0.625 pc (red) resolutions and the green squares refer to the MHD run with
1.25 pc resolution.\emph{Right:} Fractal dimension $D$ of the most dissipative
structures (derived from eq.~\ref{scalings}) as function of the exponent
$\zeta(p)/\zeta(3)$, with $p=2,4,...,10$, shown in the left panel for the three
simulations.}
\label{slopes}
\end{figure}
Using the ESS concept we determined the velocity structure functions
$\langle \delta v_{l}^{p} \rangle$, with order $p=2,4,\cdots, 10$ as a function
of $\langle \delta v_{l}^{3}\rangle$ and their best fits slopes
$\zeta(p)/\zeta(3)$ (see Avillez \& Breitschwerdt 2006) for the simulated
interstellar gas. The slopes are displayed as function of the order $p$ in the
left panel of Figure~\ref{slopes}, where triangles (black corresponds to $\Delta
x=1.25$ pc and red to $\Delta x =0.625$ pc) and squares refer to the HD and MHD
runs, respectively.

From these slopes one can determine the Hausdorff dimension of the most
dissipative structures through the solution of (Dubrulle 1994)
\begin{equation}
\label{scalings}
\frac{\zeta(p)}{\zeta(3)}=\Theta(1-\Delta)p+\frac{\Delta}{1-\beta}
\left(1-\beta^{\Theta p}\right),
\end{equation}
where $\Delta=1-\Theta$ is a parameter depending on the Hausdorff dimension $D$
of the most dissipative structures, $\beta=1-\Delta/(3-D)$ is a measure of the
intermittency and $\Theta$ is a parameter that depends on cascade of the energy
transfer in the inertial range. If this cascade is Kolmogorov, then
$\Theta=1/3$. The above mentioned ratios correspond to $D=1.9-2.02$ (right panel
of Figure~\ref{slopes}), implying that the energy injected by supernovae into
interstellar turbulence is dissipated preferentially through shocks (i.e., 2D
surfaces).

Further analysis on the Hausdorff dimension of the most dissipative structures
can be drawn from the comparison of the simulated and theoretical predictions
(eq. \ref{scalings}) of the $\zeta(p)/\zeta(3)$ variation with p shown in
Figure~\ref{slopes} (left panel). For comparison we also show the variation
of the ratio $\zeta(p)/\zeta(3)$ observed experimentally by Benzi et al. (1993)
and predicted by the K41 and She-Lev\'eque (1994; hereafter denoted by SL94)
models in incompressible turbulence and the Burgers-Kolmogorov model (Boldyrev
2002; hereafter denoted by BK02) for supersonic turbulence. The K41 model
considers a log-normal statistics for the transfer of energy from large to small
scales and has no corrections to intermittency. The SL94 and BK02 models use a
log-Poisson statistics (see discussion in Dubrulle 1994) to describe
intermittency.

The figure shows that there exists a small discrepancy between the MHD data
(green squares) and the BK02 curve and a decrease of D with increase of p. This
variation, although within the errors of numerical noise, indicates a tendency
toward filamentary dissipative structures due to the anisotropy induced by the
magnetic field. 

\section{Discussion and Final Remarks}

The dissipation of energy in interstellar turbulence proceeds through shocks and
the variation of $\zeta(p)/\zeta(3)$ with $p$ is most consistent with
a log-Poisson model where at the inertial range the cascading of energy behaves
more like Kolmogorov. Similar scalings have been found by Boldyrev et al.
(2002) and Padoan et al. (2004) for the dissipation of energy in molecular
clouds. Note, however, that our simulations include molecular clouds and its
fragmentation and cover a larger range of scales (from kpc to 0.6 pc) than those
used by these authors. The simulations also show that only a small amount of
energy is dissipated through shocks in the inertial range and the energy decay
follows a quasi-Kolmogorov spectrum. However, the shock structures start to play
an important r\^ole in energy transfer and dissipation near the small scales.

\begin{acknowledgments}
This research is supported by Portuguese Foundation for Science and Technology
(FCT) through project POCTI/FIS/58352/2004.
\end{acknowledgments}


\begin{thebibliography}{99}

\bibitem[]{ab2004a} Avillez, M.~A., \& Breitschwerdt,~D. 2004, A\&A, 425, 899 %
\bibitem[]{ab2005a} Avillez, M.~A., \& Breitschwerdt,~D. 2005, A\&A, 436, 585 %
\bibitem[]{ab2006} Avillez, M.~A., \& Breitschwerdt,~D. 2006, Phys. Rev. Lett.,
 submitted
\bibitem[1993]{benzi93} Benzi, R., Ciliberto, S., Tripiccione, R., Baudet, C.,
Massaioli, F., \& Succi, S. 1993, Phys. Rev. E 48, R29 %
\bibitem[2002]{boldyrev2002} Boldyrev, S. 2002, ApJ 569, 841 %
\bibitem[2002]{bnp02} Boldyrev, S., Nordlund, \AA., \& Padoan, P. 2002, Phys.
Rev. Lett. 89, 31102 %
\bibitem[1994]{dubrulle94} Dubrulle, B. 1994, Phys. Rev. Lett. 73, 959 %
\bibitem[2005]{elmgree05} Elmegreen, B., \& Scalo, J. 2004, ARA\&A 42, 211 
\bibitem[]{} Haverkorn, M., Gaensler, B. M., Brown, J. C., Bizunok, N. S.,
McClure-Griffiths, N. M., Dickey, J. M., \& Green, A. J. 2006, ApJ 637, L33 
\bibitem[]{} Kaplan, S. A. 1958, in "Electromagnetic Phenomena in Cosmical
Physics", Proceedings of IAU Symposium no. 6, ed. B. Lehnert, Cambridge
University Press, p.~504
\bibitem[1941]{K41} Kolmogorov, A. N. 1941, C.R. Acad. Sci. URSS 30, 301 %
\bibitem[2004]{pjnb04} Padoan, P., Jimenez, R., Nordlund, \AA. \& Boldyrev, S.
2004, Phys. Rev. Lett. 92, 1102 %
\bibitem[1994]{SL94} She, Z.-S., \& Lev\'eque, E. 1994, Phys. Rev. Lett. 72, 336
%

\end{thebibliography}
\end{document}